\documentclass[prl, reprint, superscriptaddress, bibliography]{revtex4-1}
\usepackage[dvips]{graphicx}
\usepackage{color}
\usepackage{amsmath}
\usepackage{CJK}

\begin{document}
\begin{CJK*}{UTF8}{mj}

\title{Dichotomy between Attractive and Repulsive Tomonaga-Luttinger~Liquids in~Spin~Ladders}

\affiliation{Laboratoire National des Champs Magn\'etique Intenses, LNCMI-CNRS (UPR3228), UGA, UPS, and INSA, Bo\^{i}te Postale 166, 38042, Grenoble Cedex 9, France}
\affiliation{Neutron Scattering and Magnetism, Laboratory for Solid State Physics, ETH, Z\"urich, Switzerland}
\affiliation{Jo\v{z}ef Stefan Institute, Jamova 39, 1000 Ljubljana, Slovenia}
\affiliation{Laboratoire National des Champs Magn\'etique Intenses, LNCMI-CNRS (UPR3228), UGA, UPS, and INSA, 31400, Toulouse, France}
\affiliation{Laboratory for Quantum Magnetism, Institute of Physics, Ecole Polytechnique F\'{e}derale de Lausanne (EPFL), CH-1015 Lausanne, Switzerland}

\author{M. Jeong (정민기)}
\email{minki.jeong@gmail.com}	
\affiliation{Laboratoire National des Champs Magn\'etique Intenses, LNCMI-CNRS (UPR3228), UGA, UPS, and INSA, Bo\^{i}te Postale 166, 38042, Grenoble Cedex 9, France}
\affiliation{Laboratory for Quantum Magnetism, Institute of Physics, Ecole Polytechnique F\'{e}derale de Lausanne (EPFL), CH-1015 Lausanne, Switzerland}
\author{D. Schmidiger}
\affiliation{Neutron Scattering and Magnetism, Laboratory for Solid State Physics, ETH, Z\"urich, Switzerland}
\author{H. Mayaffre}
\affiliation{Laboratoire National des Champs Magn\'etique Intenses, LNCMI-CNRS (UPR3228), UGA, UPS, and INSA, Bo\^{i}te Postale 166, 38042, Grenoble Cedex 9, France}
\author{M. Klanj\v{s}ek}
\affiliation{Jo\v{z}ef Stefan Institute, Jamova 39, 1000 Ljubljana, Slovenia}
\author{C. Berthier}
\affiliation{Laboratoire National des Champs Magn\'etique Intenses, LNCMI-CNRS (UPR3228), UGA, UPS, and INSA, Bo\^{i}te Postale 166, 38042, Grenoble Cedex 9, France}
\author{W.~Knafo}
\affiliation{Laboratoire National des Champs Magn\'etique Intenses, LNCMI-CNRS (UPR3228), UGA, UPS, and INSA, 31400, Toulouse, France}
\author{G. Ballon}
\affiliation{Laboratoire National des Champs Magn\'etique Intenses, LNCMI-CNRS (UPR3228), UGA, UPS, and INSA, 31400, Toulouse, France}
\author{B. Vignolle}
\affiliation{Laboratoire National des Champs Magn\'etique Intenses, LNCMI-CNRS (UPR3228), UGA, UPS, and INSA, 31400, Toulouse, France}
\author{S. Kr\"amer}
\affiliation{Laboratoire National des Champs Magn\'etique Intenses, LNCMI-CNRS (UPR3228), UGA, UPS, and INSA, Bo\^{i}te Postale 166, 38042, Grenoble Cedex 9, France}
\author{A. Zheludev}
\affiliation{Neutron Scattering and Magnetism, Laboratory for Solid State Physics, ETH, Z\"urich, Switzerland}
\author{M. Horvati\'c}
\email{mladen.horvatic@lncmi.cnrs.fr}
\affiliation{Laboratoire National des Champs Magn\'etique Intenses, LNCMI-CNRS (UPR3228), UGA, UPS, and INSA, Bo\^{i}te Postale 166, 38042, Grenoble Cedex 9, France}

\begin{abstract}
We present a direct NMR method to determine whether the interactions in a Tomonaga-Luttinger liquid (TLL) state of a spin-$1/2$ Heisenberg antiferromagnetic ladder are attractive or repulsive. For the strong-leg spin ladder compound $\mathrm{(C_7H_{10}N)_2CuBr_4}$ we find that the isothermal magnetic field dependence of the NMR relaxation rate, $T_1^{-1}(H)$, displays a \emph{concave} curve between the two critical fields bounding the TLL regime. This is in sharp contrast to the \emph{convex} curve previously reported for a strong-rung ladder $\mathrm{(C_5H_{12}N)_2CuBr_4}$. We show that the concavity and the convexity of $T_1^{-1}(H)$, which is a fingerprint of spin fluctuations, directly reflect attractive and repulsive fermionic interactions in the TLL, respectively. The interaction sign is alternatively determined from an indirect method combining bulk magnetization and specific heat data.  
\end{abstract}

\maketitle
\end{CJK*}

The Tomonaga-Luttinger liquid (TLL) theory provides a universal low-energy description of one-dimensional (1d) quantum many-body systems with gapless excitations \cite{Tomonaga50PTP, Luttinger63JMP, Mattis65JMP, Haldane81PRL, Giamarchi}. The predicted original properties, such as a power-law decay of the correlation functions, have been successfully observed in experiments for a wide variety of systems spanning the organic conductors \cite{Schwartz98PRB}, carbon nanotubes \cite{Egger97PRL, Kane97PRL, Bockrath99Nat, Ishii03Nat}, semiconductor quantum wires \cite{Auslaender02Sci}, edge states of the fractional quantum Hall systems \cite{Grayson98PRL}, quantum spin Hall insulators \cite{Wu06PRL}, and magnetic insulators \cite{Dender98Th, Lake05NMat}. Remarkably, these diverse systems, irrespective of their microscopic details, can be universally described within the TLL framework by only two parameters: the renormalized velocity of excitations $u$ and the TLL parameter $K$ that characterizes the sign and the strength of the interactions \cite{Giamarchi}. Recently, direct tuning of these parameters by external means has been demonstrated for a few systems, e.g., the interaction strength can be varied by an applied magnetic field in magnetic insulators \cite{Klanjsek08PRL, Jeong13PRL, Schmidiger14ETH}, or by a gate voltage in the quantum spin Hall edge states \cite{Li15PRL}. This has opened the way for a full control and exploitation of the TLL properties as well as the firm tests of the theory.

One of the recent notable achievements in this regard is the experimental realization of a TLL with \emph{attractive} interactions, i.e., $K>1$, in a magnetic insulator $\mathrm{(C_7H_{10}N)_2CuBr_4}$, called DIMPY for short \cite{Hong10PRL, Ninios12PRL, Schmidiger12PRL, Schmidiger13PRL, Jeong13PRL, Schmidiger14ETH, Povarov15PRB}. This organometallic compound features weakly-coupled (quasi-1d) ladder-like structure of the Cu$^{2+}$ ions carrying $S=1/2$ spins, which interact via the Heisenberg antiferromagnetic (AFM) superexchange \cite{Shapiro07JACS}. Generally, the ground state of a spin ladder with gapless spectrum induced by an applied magnetic field $H$, that is larger than the lower-critical field $H_{c1}$ but smaller than the upper-critical field $H_{c2}$, is described as a TLL of interacting spinless fermions \cite{Giamarchi}. The characteristics of the interactions between the fermions, parameterized by $K$, are determined by the ratio of the underlying exchange-coupling strengths along the leg and the rung, $J_\mathrm{leg}/J_\mathrm{rung}$, and further tuned by $H$ \cite{Giamarchi}. DIMPY remains a unique \emph{strong-leg} ladder compound ($J_\mathrm{leg}/J_\mathrm{rung}=1.7$) with an experimentally accessible TLL regime \cite{Jeong13PRL, Schmidiger14ETH, Shapiro07JACS, White10PRB, Hong10PRL, Schmidiger11PRB, Ninios12PRL, Schmidiger12PRL, Schmidiger13PRB, Schmidiger13PRL, Povarov15PRB}, which has long been predicted to have $K>1$ \cite{Giamarchi99PRB, Furusaki99PRB}. On the other hand, most other quasi-1d spin-$1/2$ AFM compounds, e.g., a strong-rung ladder \cite{Klanjsek08PRL, Ruegg08PRL, Bouillot11PRB} or a spin chain \cite{Zheludev07PRB, Klanjsek15PRL, Klanjsek15PRB, Halg15PRB, Jeong15PRB}, realize repulsive TLLs, i.e., with $K<1$.

\begin{figure*}
\centering
\includegraphics[width=0.95\textwidth]{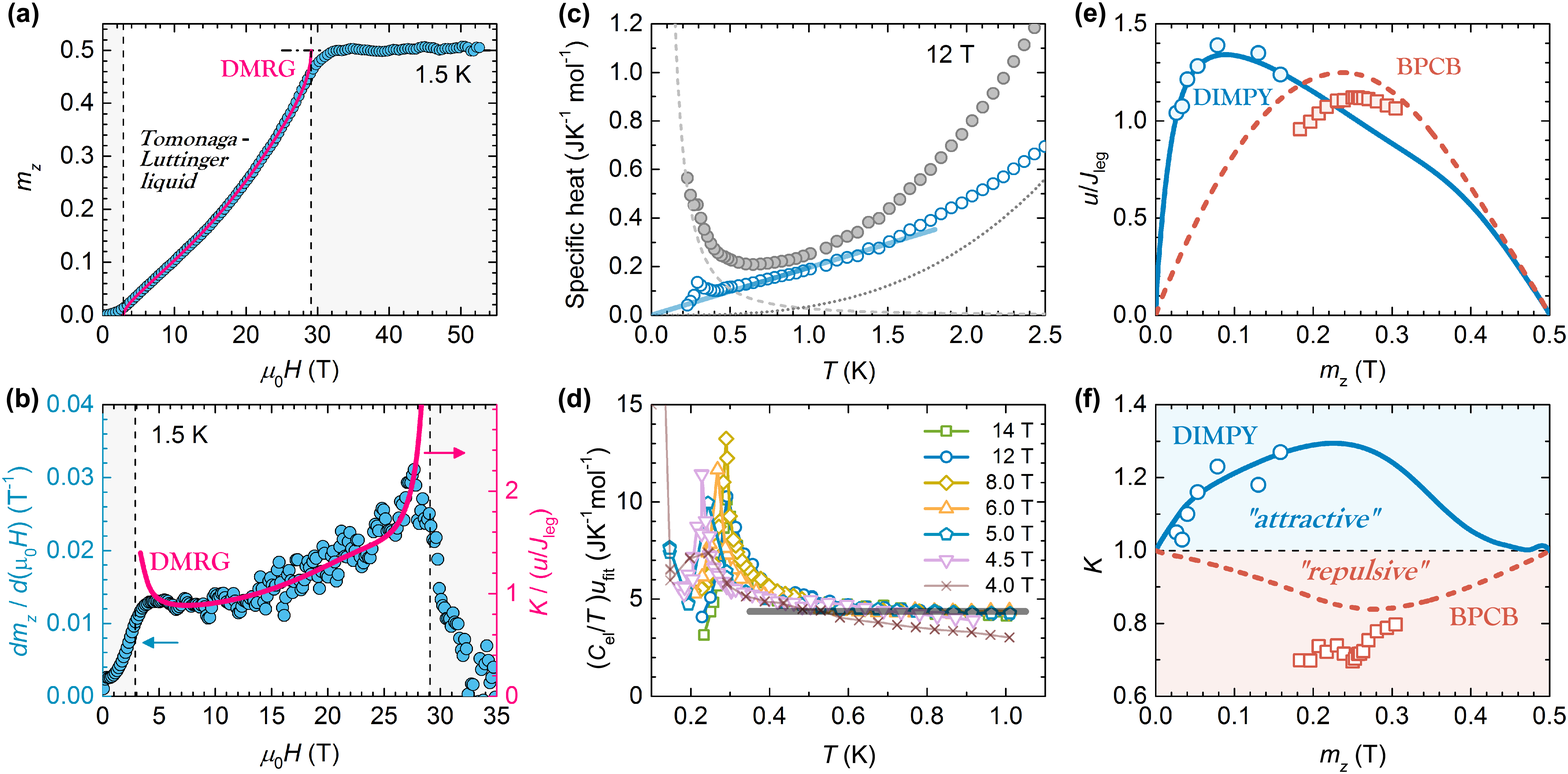}
\caption{(a) Reduced magnetization as a function of applied field, $m_z(H)$, at 1.5 K (circles), and the corresponding DMRG calculation (solid line) taken from Ref.~\cite{Schmidiger12PRL}. Dashed lines mark the critical fields, and dash-dotted line the expected saturation value. (b) Field derivative of the reduced magnetization, $\partial m_z/\partial (\mu_0H)$, as a function of $H$ (circles, left axis). The corresponding $K/(u/J_\mathrm{leg})$ from numerical calculation using $J_\mathrm{leg}=16.5$ K \cite{Schmidiger12PRL} are overlaid as a solid line (right axis). (c) Specific heat $C$ as a function of temperature for 12 T (filled circles). Dashed line represents the nuclear Schottky contribution and dotted line the lattice contribution. Open circles are electronic (magnetic) specific heat $C_\mathrm{el}$ after subtracting the nuclear and lattice contributions. Solid line identifies the $T$-linear TLL regime. (d) $C_\mathrm{el}/T$ as a function of temperature in different fields, normalized by the $u_\mathrm{fit}$ derived from the linear fit to $C_\mathrm{el}$ (see the text). (e) $u/J_\mathrm{leg}$ as a function of $m_z$ deduced from the specific heat analysis (circles). The corresponding values for BPCB (squares) obtained by analyzing the data reported in Ref.~\cite{Ruegg08PRL} are also shown. DMRG results for DIMPY~\cite{Schmidiger12PRL} and BPCB~\cite{Bouillot11PRB, Schmidiger12PRL} are plotted by solid and dashed lines, respectively. (f) $K$ as a function of $m_z$ deduced from the combined analysis of the specific heat and magnetization data~\cite{Klanjsek08PRL}.}\label{Fig1}
\end{figure*}

A direct experimental evidence for the attractive interactions in DIMPY was first obtained from the nuclear magnetic resonance (NMR) spin-lattice relaxation rate $T_1^{-1}$ measurements \cite{Jeong13PRL}. $T_1^{-1}$ as a function of temperature $T$ was shown to follow a theoretically-predicted form $T_1^{-1}\propto T^{1/2K-1}$ where $K$ varies between 1 and 2 as a function of $H$. However, the obtained $K$ values in the measured 3.5 T $<\mu_0H<$ 15 T range, where $\mu_0H_{c1}\simeq 2.6$ T \cite{White10PRB, Hong10PRL, Ninios12PRL, Schmidiger12PRL, Jeong13PRL} and $\mu_0H_{c2}\simeq 29$ T \cite{White10PRB, Schmidiger12PRL}, were larger by up to 60 \% than those calculated using the density matrix renormalization group (DMRG) method \cite{Schmidiger12PRL}. This quantitative discrepancy was tentatively attributed to the effects of (weak) 3d exchange couplings \cite{Jeong13PRL}, which have not been included in purely 1d theoretical expressions \cite{Klanjsek08PRL, Bouillot11PRB}, and remain to be investigated. Moreover, due to a relatively weak quantitative variation of the power-law exponent $1/2K-1$ for the $K$ values lager than 1, combined with a limited $T$ window of the power-law behavior, an unambiguous determination whether the interactions are attractive or repulsive becomes a demanding task.

Based on our experimental data for DIMPY, we demonstrate in this Letter that the isothermal $T_1^{-1}(H)$ is a reliable observable that \emph{directly} and \emph{qualitatively} distinguishes between the attractive and the repulsive TLLs. Prior to this, we also present thermodynamic evidence for the attractive interactions in DIMPY by combining bulk magnetization and specific heat data, which is shown, however, to be indirect and vulnerable to errors, in contrast to the method based on NMR.

All measurements were done on single crystal samples \cite{Yankova12PM, Schmidiger14ETH}. Magnetization was measured using the compensated coil technique in a pumped $^4$He cryostat and a pulsed field up to 52 T. The 14 MJ capacitor bank at LNCMI, Toulouse, was used to generate the pulsed field with a typical rise and fall time of 30 ms and 120 ms, respectively. Specific heat was measured using the relaxation technique with a Physical Property Measurement System (PPMS) equipped with a $^3$He-$^4$He dilution refrigerator and a 14 T magnet. High-field $^{14}$N NMR measurements were performed in the $15-30$ T field range using a 20 MW resistive magnet at LNCMI, Grenoble. The lower-field NMR data were obtained using a 15 T superconducting magnet. $T_1^{-1}$ values were obtained by the saturation-recovery method using a standard pulsed spin-echo technique \cite{Jeong13PRL}.

Figure \ref{Fig1}(a) shows the reduced magnetization curve $m_z(H)$ at 1.5 K in a field along the crystallographic $a$ axis (ladder direction). The measured magnetization values are normalized by the expected saturation magnetic moment, such that $m_z$ refers to the longitudianl spin expectation value $\langle S_z \rangle$ per Cu$^{2+}$ ion. Up to the closing of the spin gap at $\simeq 2.5$ T, there is only a minor increase of $m_z$ with $H$ \cite{White10PRB, Hong10PRL, Ninios12PRL, Schmidiger12PRL}. This is then followed by a progressively steeper increase of $m_z$ in the gapless phase up to $\sim 30$ T, where a gapped fully-polarized phase is reached and $m_z$ levels off. The measured $m_z(H)$ is in excellent agreement with the previous DMRG calculation \cite{Schmidiger12PRL} except close to the critical fields where finite-temperature effects are expected. The saturation field $H_{c2}$ is estimated to be $30(1)$ T. The magnetization curve up to 40 T was previously reported using a similar pulsed-field measurement, but for the field along the $b$ axis \cite{White10PRB}. We note, however, that the corresponding $m_z(H)$ dataset taken at 1.6 K is significantly rounded off and scattered, particularly upon approaching the saturation, which would hinder its application for deriving the TLL properties. On the other hand, our 1.5 K dataset from the TLL regime shows a well-defined $m_z(H)$ profile and levels off sharply, similar to the 480 mK data in Ref.~\cite{White10PRB}.

The actual raw response to the pulsed magnetic field in the $m_z(H)$ measurements is given by a differential magnetization with respect to the field, i.e., $\partial m_z/\partial (\mu_0H)$, which is presented in Fig.~\ref{Fig1}(b). This observable has a simple relation with the TLL parameters as 
\begin{equation}\label{eq:mag}
\frac{\partial m_z}{\partial \left(\mu_0H\right)}=\frac{g\mu_\mathrm{B}}{2\pi k_\mathrm{B}}\left(\frac{K}{u}\right),
\end{equation}
where $\mu_0$ is the vacuum permeability, $\mu_\mathrm{B}$ the Bohr magneton, and $u$ is in kelvin units \cite{Giamarchi, Bouillot11PRB, Schmidiger14ETH}. We used the previously calculated $K$ and $u$ values \cite{Schmidiger12PRL} to obtain the right-hand side of Eq.~\eqref{eq:mag}. The results plotted in Fig.~\ref{Fig1}(b) are in excellent agreement with the $\partial m_z/\partial (\mu_0H)$ dataset.

The molar specific heat $C$ was measured as a function of temperature in different magnetic fields. The representative 12 T data (filled circles) are shown in Fig.~\ref{Fig1}(c). As the temperature decreases, $C$ decreases until it reaches a minimum, and then increases rapidly by nuclear Schottky anomaly. The overall $C(T)$ behavior remains the same for other field values between 3 T and 14 T, in the gapless region. The electronic (magnetic) part, $C_\mathrm{el}$, was obtained by subtracting the nuclear spin and lattice contributions from the raw data \cite{SM}. A $T$-linear regime characteristic of a TLL \cite{Giamarchi} is identified below 1 K, while a small peak appearing close to 300 mK corresponds to the magnetic ordering transition \cite{Schmidiger12PRL, Ninios12PRL, Jeong13PRL}.

The molar $C_\mathrm{el}$ for a TLL is known to be inversely proportional to $u$ as 
\begin{equation}\label{eq:Cp}
C_\mathrm{TLL} = N_\mathrm{A}\frac{\pi k_\mathrm{B}T}{6u},
\end{equation}
where $N_\mathrm{A}$ is the Avogadro constant \cite{Giamarchi, Bouillot11PRB, Schmidiger12PRL, Schmidiger14ETH}. We apply Eq.~\eqref{eq:Cp} to the $T$-linear regime of the $C_\mathrm{el}(T)$ data for different fields to obtain the $u$ value as the fitting parameter. The resulting plots of $\left(C_\mathrm{el}/T\right)u_\mathrm{fit}$ for various fields (4.5, 6, and 8 T data from Ref.~\cite{Schmidiger12PRL}) are shown in Fig.~\ref{Fig1}(d)~\cite{footnote}. The data collapsing on a flat line above the corresponding transition temperatures demonstrate the presence of the TLL regime. The obtained $u$ values are then plotted as a function of $m_z$ in Fig.~\ref{Fig1}(e), which shows a nice agreement with the DMRG calculation \cite{Schmidiger12PRL}. For comparison, we make a similar analysis of the reported specific heat data for the prototypical strong-rung ladder compound $\mathrm{(C_5H_{12}N)_2CuBr_4}$, known as BPCB \cite{Ruegg08PRL}. The obtained experimental points in Fig.~\ref{Fig1}(e) show a slight systematic deviation from the calculation by DMRG \cite{Klanjsek08PRL, Bouillot11PRB, Schmidiger12PRL}, which may be attributed to incomplete subtraction while obtaining the corresponding $C_\mathrm{el}$.

Eventually, we deduce the experimental $K$ values by combining the presented $C_\mathrm{el}$ and $\partial m_z/\partial (\mu_0H)$ data in the Wilson ratio, directly proportional to $K$ \cite{Ninios12PRL}. The obtained $K$ values are plotted as a function of $m_z$ in Fig.~\ref{Fig1}(f). For DIMPY, $K$ is found to lie in the $K>1$ range and increases with the field up to 14 T which corresponds to $m_z \simeq 0.16$. This provides a thermodynamic evidence for the attractive interactions and their field variation. The experimental data are in good agreement with the DMRG prediction taken from Ref.~\cite{Schmidiger12PRL}. A similar analysis of previous measurements, albeit limited to a narrower field range, resulted in the $K$ values that unexpectedly extend over both the attractive and the repulsive regimes~\cite{Ninios12PRL}. For comparison, we also obtain the experimental $K$ values for BPCB using the reported data \cite{Ruegg08PRL, Klanjsek08PRL}, which fall into the repulsive regime yet exhibit subtantial deviation from the calculations \cite{ Klanjsek08PRL, Bouillot11PRB} [Fig.~\ref{Fig1}(f)]. Apparently, conclusions that can be drawn from the presented bulk method are crucially affected by the limited quality of the experimentally determined $C_\mathrm{el}$ and $\partial m_z/\partial (\mu_0H)$.  The limits originate from an indirect way both observables are experimentally determined.

In the following, we revisit the analysis of the NMR relaxation data to reveal an easy and direct \emph{qualitative} criterion to distinguish between the attractive and repulsive TLLs. We employ $^{14}$N nuclei for the $T_1^{-1}(H)$ measurements, instead of $^{1}$H nuclei previously used for the $T_1^{-1}(T)$ measurements \cite{Jeong13PRL}. A different choice of the probe nuclei is of practical importance. While using $^1$H nuclei results in a large signal owing to the large gyromagnetic ratio, numerous proton sites give rise to a strongly field-dependent complex shape of the NMR spectra. As much as the overlap of $^1$H lines varies with the field, measured $T_1^{-1}$ may provide a distorted, unreliable image of spin fluctuations. This is not the case for $^{14}$N nuclei, which give rise to simple and well separated spectral lines over the entire field range \cite{Jeong13PRL}.

\begin{figure}
\centering
\includegraphics[width=0.45\textwidth]{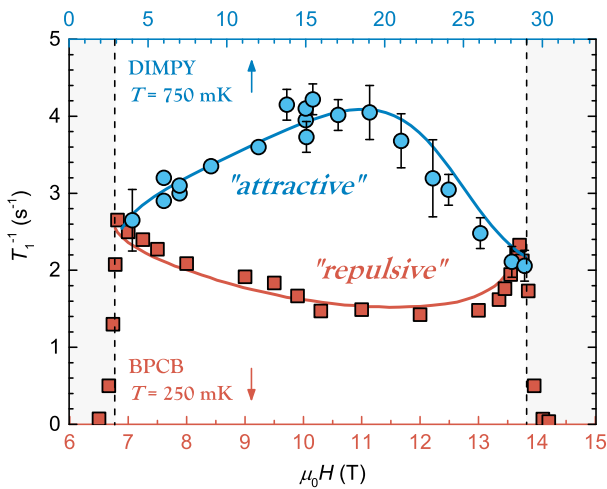}
\caption{$^{14}$N NMR relaxation rate $T_1^{-1}$ as a function of magnetic field $H$ for DIMPY at 750 mK (circles, top axis) and for BPCB at 250 mK (squares, bottom axis). The data for BPCB are reproduced from Ref.~\cite{Klanjsek08PRL}. Lines are plots of the theoretical TLL expression Eq.~\eqref{eq:T1} using the parameters calculated by DMRG \cite{Schmidiger12PRL, Klanjsek08PRL, Bouillot11PRB}.}\label{Fig2}
\end{figure}

Figure~\ref{Fig2} shows the field dependence of $^{14}$N $T_1^{-1}$ in DIMPY at a constant $T=750$ mK. $T_1^{-1}$ increases with $H$ from $H_{c1}$, displays a broad maximum, and then decreases as $H$ approaches $H_{c2}$: the $T_1^{-1}(H)$ displays an overall concave shape between $H_{c1}$ and $H_{c2}$. This is in sharp contrast to the convex shape previously reported for BPCB \cite{Klanjsek08PRL}, also shown in Fig.~\ref{Fig2}. The selected temperatures in both cases are more than two times higher than the maximum ordering temperatures of $330$~mK and $110$~mK for DIMPY and BPCB, respectively, which ensures that the systems are deep in the TLL phases.

The NMR $T_1^{-1}$ probes local spin-spin correlations in the low-energy limit. For the spin ladder in its TLL phase, the transverse correlations at $Q=\pi$ are known to be dominant \cite{Giamarchi99PRB}, and they lead to the following expression, 
\begin{equation}\label{eq:T1}
T_1^{-1} \!=\! \frac{\hbar\gamma^2A_\perp^2A_0^x}{k_\mathrm{B}u}\!\cos\!\left(\frac{\pi}{4K}\!\right)\!B\!\left(\!\frac{1}{4K},1\!-\!\frac{1}{2K}\!\right)\!\!\left(\!\frac{2\pi T}{u}\!\right)^{\!\frac{1}{2K}-1},
\end{equation}
where $\gamma$ is the nuclear gyromagnetic ratio, $A_\perp$ the transverse hyperfine coupling constant, $A_0^x$ the amplitude of the correlation function, and $B(x,y)=\Gamma(x)\times\Gamma(y)/\Gamma(x+y)$ \cite{Klanjsek08PRL, Bouillot11PRB}. As the field variations of $K$, $u$, and $A_0^x$ are calculated by DMRG \cite{Schmidiger12PRL, Klanjsek08PRL}, Eq.~\eqref{eq:T1} allows direct comparison between the experimental and the numerical results using an overall scale factor $A_\perp$ as a fitting parameter. Lines in Fig.~\ref{Fig2} are thus obtained $T_1^{-1}(H)$ curves for DIMPY as well as for the previously reported BPCB \cite{Klanjsek08PRL}, which nicely match with the experimental data; they indeed reproduce perfectly the contrasting concave and convex curves.

We further find that these two contrasting behaviors are generic, and propose them as a hallmark to distinguish between the repulsive and attractive interactions in the TLL spin systems. This can be readily demonstrated by first considering the vicinity of $H_{c2}$, where the following analytical expressions are known for the asymptotic behaviors: $A_0^x \propto \sqrt{1/2 - m_z}$ \cite{Hikihara04PRB} and $u \propto (1/2 - m_z)$, where $(1/2 - m_z) \propto \sqrt{H_{c2} - H}$. Knowing also that the linear development $\cos\left(\pi/4K\right)B\left(1/{4K}, 1-1/2K\right) \simeq 4.20 (K-1) + 3.71$ is remarkably precise, within 1\% error in the whole $0.62 < K < 2$ range, we get from Eq.~\eqref{eq:T1},
\begin{equation}\label{eq:T1ofK}
T_1^{-1}(H\rightarrow H_{c2}) \propto (H_{c2} - H)^{\frac{K-1}{4}} \left(4.20(K - 1) + 3.71\right),
\end{equation}
to the first order in ($K-1$). Both terms on the right hand side of Eq.~\eqref{eq:T1ofK} increase (decrease) when $K$ departs from 1 towards larger (smaller) values by decreasing $H$ below $H_{c2}$ . The same behavior is expected close to $H_{c1}$, although the $A_0^x(m_z)$ and $u(m_z)$ dependences are renormalized as compared to their $H_{c2}$ counterparts given above. Connecting these two asymptotic behaviors in a smooth way then provides the proof that Eq.~\eqref{eq:T1} indeed predicts either the concave or the convex $T_1^{-1}(H)$ curve, reflecting directly the sign of interactions over the field range between $H_{c1}$ and $H_{c2}$. The same argument can be used when the longitudinal spin fluctuations are dominant, as is the case in BaCo$_2$V$_2$O$_8$ \cite{Klanjsek15PRB}. This occurs for $K<1/2$, which represents the {\em strongly repulsive} case \cite{Okunishi07PRB}. Eq.~\eqref{eq:T1} valid for the transverse spin fluctuations can be applied also to the case of longitudinal spin fluctuations by replacing $1/(2K) \longrightarrow 2K$ \cite{Klanjsek15PRB, Okunishi07PRB}. Taking into account the corresponding amplitude of the correlation function $A_1^z$ \cite{Hikihara04PRB}, also in this case the shape of $T_1^{-1}(H)$ turns out to be convex, thus establishing a full relation between the repulsive case and the convex shape.

We note that the concave/convex shape of the $K(H)$ curve has already been predicted theoretically for the spin ladder systems \cite{Hikihara01PRB}, but there has been no clear experimental demonstration thus far. Here we discover that $T_1^{-1}(H)$ closely follows the $K(H)$ behavior and serves as the first \emph{direct} probe of the interaction sign, without complications and errors that might easily occur during many steps of the (indirect) analysis of the bulk property measurements.

There are systems other than spin ladders where a crossover between the attractive and repulsive regimes can be explored by tuning a model parameter \cite{Haga00JPSJ, Hikihara01PRB, Sato09PRB, Sato11PRB}. For instance, a bond-alternating AFM-FM spin-$1/2$ chain is among a few feasible spin models known to support an attractive TLL, while the AFM-AFM chain supports a repulsive one \cite{Sakai95JPSJ}. The former model has been recently realized in an organic magnet \cite{Yamaguchi15PRB}. A field-induced TLL of the Haldane (i.e., spin-1) chain also supports attractive interactions \cite{Konik02PRB}, while adding a large enough single-ion anisotropy to this system induces a crossover into a repulsive regime, as expected in an organometallic $\mathrm{NiCl_2\cdot4SC(NH_2)_2}$ \cite{PaduanFilho04PRB, Mukhopadhyay12PRL}. These systems thus offer an opportunity to use our NMR criterion to reveal the nature of their interactions. Indeed, the reported $1/T_1$ data \cite{Goto06PRB} for a TLL state in a Haldane chain $\mathrm{(CH_3)_4NNi(NO_2)_3}$ are consistent with our prediction. The quantum Hall edge states are yet another alternative physical system where both repulsive and attractive TLLs have been realized \cite{Grayson07PRB}. We hope that our results will stimulate a similar line of efforts to characterize the interactions in other realizations of TLLs.

To conclude, isothermal NMR $T_1^{-1}(H)$ is established as a sensitive probe of the sign of the interactions between the spinless fermions in the TLL state of a spin-$1/2$ Heisenberg AFM ladder. $T_1^{-1}(H)$ displays a concave curve for the attractive TLL of a strong-leg ladder DIMPY, which is in sharp contrast to the convex one for the repulsive TLL of a strong-rung ladder BPCB. This experimental finding is well reproduced by the theoretical TLL expression for $T_1^{-1}(H)$ combined with previous DMRG calculations. Our results thus establish a direct experimental criterion that qualitatively distinguishes between the attractive and repulsive TLLs. Furthermore, combined bulk magnetization and specific heat data are shown to provide an alternative, thermodynamic evidence for an attractive TLL in DIMPY.

We thank T. Giamarchi for discussions, P. Bouillot for sharing the DMRG data, and P. Babkevich, M. Grbi\'c, and I. \v{Z}ivkovi\'c for useful comments on the manuscript. This work was supported by the European Commission contract EuroMagnet II (No. 228043), the French ANR project BOLODISS, and the Swiss National Science Foundation, Division II. M.J. is grateful to support by European Commission through Marie Sk{\l}odowska-Curie Action COFUND (EPFL Fellows) and European Research Council grant CONQUEST.

\bibliography{bib_dichotomy}

\begin{thebibliography}{53}
\bibitem [{foo()}]{footnote}%
  \BibitemOpen
\bibfield  {journal} {  }\href@noop {} {\bibinfo  {journal} {The gradual
  deviation of the 4 T dataset with increasing temperature is attributed to the
  proximity of $H_{c1}$ and the corresponding quantum critical region. The data
  fitting is performed in the 0.6 K $<T<$ 1 K range for $H\geq$ 5 T, and in the
  0.4 K $<T<$ 0.6 K range for $H<$ 5 T.}\ }\BibitemShut {NoStop}%
\end{thebibliography}

\end{document}